\documentstyle[epsfig,color,pslatex]{mn}

\definecolor{green}{rgb}{0,0.7,0}
\definecolor{blue}{rgb}{0,0,0.7}
\definecolor{red}{rgb}{0.7,0,0}
\definecolor{magenta}{cmyk}{0.1,0.8,0,0.1}
\definecolor{mauve}{rgb}{0.5,0,0.5}

\begin{document}

\title[Annihilation and radio emission of galactic jet sources]{Pair annihilation and radio emission from galactic jet sources: The case of Nova Muscae}

\author[C.R. Kaiser \& D.C. Hannikainen]{Christian R. Kaiser\thanks{email: crk@astro.soton.ac.uk} and Diana C. Hannikainen\\
Department of Physics \& Astronomy, University of Southampton, Southampton SO17 1BJ}

\maketitle

\begin{abstract}
In the hard X-ray spectra of some X-ray binaries line features at
around 500~keV are detected. We interpret these as arising from pair
annihilation in relativistic outflows leading to a significant Doppler
shift of the lines' frequencies. We show how this can be used to
accurately determine the bulk velocity and orientation to the line of
sight of the outflows. Constraints on the energy requirements of such
outflows are also derived. Furthermore, we show that a small fraction
of pairs escaping the annihilation region may give rise to the radio
synchrotron emission observed in some of these objects. We apply these
ideas to the hard X-ray and radio observations of Nova Muscae 1991. In
this object, the energy requirements seem to rule out a large proton
fraction in the outflows.
\end{abstract}


\section{Introduction}

It has been known for some time now that a handful of
Galactic X-ray transients has exhibited radio jets. 
The first X-ray binary for which a 
kinematic model involving radio jets was proposed to 
explain its bizarre observed spectrum 
was SS~433 in 1979 (Abell \& Margon 1979). 
Subsequently, in 1994, apparent superluminal motion of ejected material 
was observed for the first time in our Galaxy from 
GRS~1915$+$105 (Mirabel \& Rodr{\'\i}guez 1994\nocite{mr94}). 
The velocity of the ejecta was estimated to be 0.92$c$ for 
an assumed distance of 12.5~kpc. 
Since that first discovery, four other {\it transients}
have shown apparent superluminal motion: GRO~J1655$-$40  (Hjellming \&
Rupen 1994; Tingay et al. 1995); XTE~J1748$-$288 (Rupen, Hjellming \&
Mioduszewski 1998; V4641~Sgr (Hjellming et al. 2000); and XTE~J1550$-$564
(Hannikainen et al. 2001).

In addition to the transient jet sources, there has also been evidence
of steady compact radio jets during the low hard state of Galactic
black hole candidate X-ray binaries from, 
for example, Cyg~X-1 (Stirling et al. 2001)
and 1E~1740.7$-$2942, or ``The Great Annihilator'' (e.g. Mirabel et
al. 1992; see Fender 2001 for a full review on
radio jets from X-ray binaries in the low hard state).
The presence of jets in X-ray
binaries gives rise to radio emission and
in this paper we will take the observation of radio emission to
indicate the presence of jets.  Note however, that in the absence of
resolved radio observations, this emission may also stem from other
regions within the source.

The composition of the jets, either electron-positron or
electron-proton plasma, is still not fully established.  For example,
Gliozzi, Bodo \& Ghisellini (1999) investigated the role of
electron-positron pairs (amongst other possibilities) as energy
carriers from the inner jet regions, and have excluded a pair plasma
as a viable possibility. They argue that either hot or cold pairs
cannot survive the annihilation.

In this paper we propose that the annihilation line features observed
in the hard X-ray spectra of some X-ray transients 
arise from pairs in a bipolar outflow. 
At the time of annihilation, this outflow is
already accelerated to relativistic bulk speeds causing a significant
Doppler shift of the frequency of the annihilation lines. We also show
that the subsequent emission of radio synchrotron radiation from the
outflow may be caused by only a small fraction of pairs escaping from
the annihilation region. We apply this idea to radio and hard X-ray
observations of Nova Muscae 1991, thus constraining the properties of
the possible outflow from this system. The energy requirements for
this source rule out a large contribution of protons to the outflow.
We would like to point out that 
1E~1740.7$-$2942 has exhibited annihilation features near 511~keV 
(e.g. Mandrou et al. 1990), and that there is some evidence of 
the same in Cyg~X-1 (Ling \& Wheaton 1989). We are currently working
on expanding the arguments presented in this paper to encompass the
other two sources (Hannikainen \& Kaiser, in preparation).

In Section \ref{sec:dop} we discuss under which plasma conditions we
can expect Doppler-shifted annihilation lines. Section \ref{sec:sync}
briefly reviews radio synchrotron emission. The case of Nova Muscae
1991 is investigated in Section \ref{sec:nm} and in Section
\ref{sec:conc} we summarise our results.

\section{Doppler-shifted annihilation lines}
\label{sec:dop}

\subsection{Conditions for a strong, narrow annihilation line}

The direct annihilation of an electron-positron pair results in the
production of two $\gamma$-ray photons, each with an energy of
511\,keV in the rest-frame of the annihilating particles. In the case
of a hot pair plasma the resulting emission line is broadened by the
random thermal motion of the pairs (Ramaty \& M{\'e}sz{\'a}ros
1981\nocite{rm81}). Also, the annihilation of a thermal pair plasma is
accompanied by bremsstrahlung from the pairs. The bremsstrahlung
emissivity around 511\,keV exceeds that of the annihilation process at
temperatures above $\Theta = k T / (m_{\rm e} c^2)=3$ (Svensson
1982\nocite{rs82}; Macio\l ek-Nied{\'z}wiecki, Zdziarski \& Coppi
1995\nocite{mzc95}, hereafter MZC95). Below a temperature of about
$10^6$\,K the annihilation proceeds via the formation of positronium
leading to a distinctive low-energy wing of the annihilation line
(e.g. Longair 1994\nocite{ml94}). Thus in the case of a thermal pair
plasma the plasma temperature must lie roughly in the range $10^6$\,K
to $4\times 10^9$\,K for a narrow annihilation line to be observable.

In a plasma, pairs may be produced by various mechanisms.
Unless there is a strong external radiation field
containing many photons with energies in excess of 511\,keV, all
of these require the presence of highly energetic electrons and
soft seed photons. The electrons upscatter the soft photons beyond
the pair creation threshold and thus the plasma starts producing
pairs. As a by-product of the pair production, a strong
Comptonization spectrum is emitted by the plasma. MZC95 point out
that this Comptonization spectrum completely swamps the
annihilation line in thermal plasmas. An observable narrow line
can only be produced if a large fraction of the pairs escapes from
the plasma and has time to cool before annihilating. However, this
implies the existence of a rather sharp boundary between a fairly
hot pair plasma bathed in an intense radiation field of soft
photons and a region virtually devoid of any soft photons. Also,
although the pairs escape the plasma, the volume containing the
pairs must be confined at least for a time longer than the cooling
time. Any expansion would lead to enormously reduced annihilation
rates. Although it is certainly possible to construct such
geometries, the existence of non-thermal leptons within the plasma
allows for a simpler way of producing a narrow annihilation line.

In non-thermal plasmas relativistic electrons or pairs may be injected
into the plasma. The injected pairs and those produced in the plasma
may cool to sub-relativistic energies and thermalize before
annihilating, thus leading to a narrow annihilation line (Lightman \&
Zdziarski 1987\nocite{lz87}, hereafter LZ87). If the line is strong,
it can rise above the Comptonization spectrum and becomes
detectable. This requires a high pair yield, $Y$, defined as the ratio
of the energy converted to pairs and the energy supplied to the
plasma. The highest pair yields can be achieved when the plasma is
`photon-starved', i.e. when the number of injected relativistic
photons strongly exceeds that of the injected soft photons (Zdziarski,
Coppi \& Lamb 1990\nocite{zcl90}). In this case, $Y\sim 0.25$ and a
strong, narrow annihilation line above the Comptonization continuum
becomes observable.

We conclude that the observation of a narrow annihilation line most
likely indicates a plasma with strong injection of non-thermal
electrons or pairs. The injection of leptons into a spherical volume
of radius $R$ is characterised by the compactness (e.g. LZ87)

\begin{equation}
l_{\rm e} = \frac{L_{\rm e} \sigma _{\rm T}}{R m_{\rm e} c^3},
\label{compact}
\end{equation}

\noindent where $\sigma _{\rm T}$ is the Thomson cross-section and
$L_{\rm e}$ is the power of the electron injection,

\begin{equation}
L_{\rm e} = \frac{4 \pi R^3}{3} m_{\rm e} c^2 \int Q_{\rm l}(\gamma)
\left( \gamma -1 \right) \, d\gamma.
\label{luminosity}
\end{equation}

\noindent Here, $Q_{\rm l}(\gamma)$ is the rate of injection of
leptons with Lorentz factor $\gamma$ per unit volume per unit time per
unit $\gamma$. If the compactness $l_{\rm e}$ can be inferred from the
observations of an annihilation line, then Equations~(\ref{compact})
and (\ref{luminosity}) can be used to constrain $Q_{\rm l}(\gamma)$.

\subsection{Relativistic Doppler-shifts}

\begin{figure}
\begin{center}
\epsfig{file=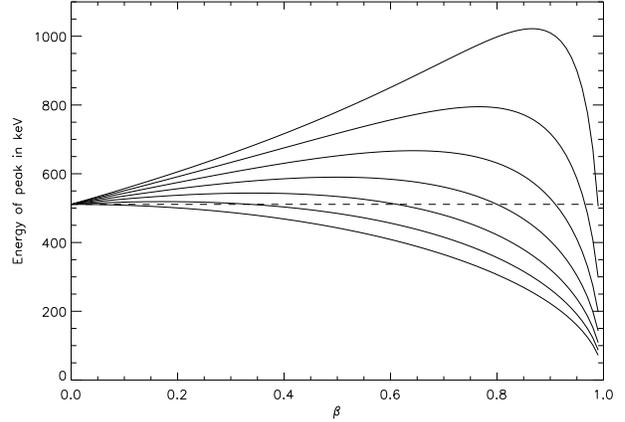, width=8.45cm}
\caption{Doppler-shift of positron annihilation radiation from
relativistically moving material approaching an observer. The dashed
line shows the unshifted rest-frame energy of the annihilation line at
511\,keV. The solid lines show the energy of the line as function of
the velocity of the material as observed at various angles to the line
of sight (Equation~\ref{delta}). $\theta$ is increasing in steps of
$10^{\circ}$ from $30^{\circ}$ (top) to $90^{\circ}$ (bottom).}
\label{fig:doppler}
\end{center}
\end{figure}

Any emission of relativistically moving material is Doppler-shifted in
its frequency. For material moving with bulk velocity $v_{\rm b}=\beta
c$ at an angle $\theta$ to the line of sight to the observer, the
observed frequency, $\nu$, of radiation emitted at frequency $\nu'$ in
the rest-frame of the material is given as

\begin{equation}
\nu = \frac{\nu'}{\gamma _{\rm b} \left( 1 \pm \beta \cos \theta
\right)} = \nu' \delta_{\pm},
\label{delta}
\end{equation}

\noindent where $\gamma _{\rm b}$ is the Lorentz factor corresponding
to the velocity $\beta$ (e.g. Rybicki \& Lightman
1979\nocite{rl79}). The upper signs correspond to material receding
along the line of sight to the observer while the lower signs indicate
approaching material. From Equation~(\ref{delta}) it is
clear that radiation of material receding from an observer is always
redshifted. However, for approaching material the emission may be
blueshifted or redshifted, depending on the combination of $\beta$ and
$\theta$. Figure \ref{fig:doppler} shows this effect for the example
of positron annihilation radiation emitted at a rest-frame energy of
511\,keV. Any combination of $\beta$ and $\theta$ below the dashed
line in this figure corresponds to redshifted emission from the
approaching material. The range of velocities which result in such a
Doppler redshift is largest for angles to the line of sight close to
$90^{\circ}$.

For a source with a bipolar outflow with components travelling in
opposite directions there will always be one approaching and one
receding component.  If both components emit a strong annihilation
line, we expect to observe two such lines at frequencies $\nu_{-}$ and
$\nu_{+}$. Assuming the same bulk velocity for both components,
Equation~(\ref{delta}) can then be solved for the bulk velocity of the
components,

\begin{equation}
\gamma_{\rm b} = \frac{\nu'}{2} \left( \frac{1}{\nu _{-}} +
\frac{1}{\nu _{+}} \right),
\label{gamma}
\end{equation}

\noindent and for the angle of the bulk velocity to the line of sight

\begin{equation}
\cos \theta = \frac{1}{\beta} \left( \pm \frac{\nu'}{\gamma _{\rm b}
\nu _{\pm}} \mp 1 \right),
\label{cos}
\end{equation}

\noindent where again the upper signs correspond to the receding
component and the lower signs to the approaching one.

In practice it may be difficult to observe the annihilation line from
the receding component as this will be strongly deboosted. In fact,
the ratio of the line fluxes provides for an independent check on the
quantity $\beta \cos \theta$ as (Rybicki \& Lightman 1979\nocite{rl79})

\begin{equation}
\frac{F_{-}}{F_{+}} = \left( \frac{1+\beta \cos \theta}{1-\beta \cos
\theta} \right) ^k,
\label{fluxratio}
\end{equation}

\noindent with $k=3$ for discrete components and $k=2$ for continuous
jets. It is usually difficult to determine the line fluxes accurately
from observations. This, combined with the ambiguity of the value of
$k$, implies that the line flux ratio is of limited use.

\section{Radio synchrotron emission and absorption}
\label{sec:sync}

Synchrotron emission arises from relativistic electrons and/or
positrons spiralling in a magnetic field. It is usually described by a
power law spectrum with $L_{\nu} \propto \nu^{\alpha}$. Both the
relativistic particles and the magnetic field store energy, but how
the total energy is divided between the two cannot be determined from
the observed radiation alone. However, it is well known that there is
a minimum total energy required to produce a given monochromatic
synchrotron luminosity, $L_{\nu}$. This is usually expressed in terms
of the magnetic field corresponding to this condition

\begin{equation}
B_{\rm min} = \left[ \frac{3 \mu _0 G(\alpha) L_{\nu}}{2 V}
\right]^{2/7},
\label{bmin}
\end{equation}

\noindent where $V$ is the volume of the emitting region and
$G(\alpha)$ is a numerical constant depending on the slope of the
synchrotron spectrum, $\alpha$, and on $\nu _{\rm min}$ and
$\nu_{\rm max}$, the limits of the spectrum. It is
usually assumed that $\nu_{\rm min} \sim 10$\,MHz and $\nu_{\rm
max} \sim 100$\,GHz. Note here that the exact choice for the
spectral cut-offs, and particularly the choice for $\nu_{\rm
max}$, has very little influence on the value of $G(\alpha)$ for
$\alpha \sim 0.5 \rightarrow 0.9$, typical for optically thin
synchrotron radiation (e.g. Longair 1994\nocite{ml94}). The
energy in the relativistic particles is then given by

\begin{equation}
W_{\rm part} = \frac{2 V B_{\rm min}^2}{3 \mu_0} = V m_{\rm e} c^2
\int P(\gamma) (\gamma -1) \, d\gamma,
\label{wpart}
\end{equation}

\noindent where $P(\gamma)$ is the number of relativistic particles
with Lorentz factor $\gamma$ per unit volume per unit $\gamma$. For a
given synchrotron luminosity and a known volume of the emission region
we can therefore constrain $P(\gamma)$. Obviously, this is just an
estimate as the distribution of the energy between the relativistic
particles and the magnetic field may depart strongly from the minimum
energy values. However, such a departure would lead to much higher
overall energy requirements.

The estimates presented above implicitly assume that the emission
region is optically thin to radio waves. This is not necessarily the
case, particularly in the inner, dense regions of the source. For
$P(\gamma) = P_0 \gamma ^{-p}$ the optical depth is

\begin{equation}
\tau = F(p) n_0 B_{\rm min}^{(p+2)/2} \nu'^{-(p+4)/2} R,
\label{depth}
\end{equation}

\noindent where $F(p)$ is a function of $p$ only (e.g. Longair
1994\nocite{ml94}) and $R$ is the typical path length photons have to
travel through the emission region. If the magnetic field is tangled
on scales significantly smaller than the physical size of the emission
region, its energy density obeys $u_{\rm mag} \propto V^{-4/3}$. So
for a spherical volume of radius $R$ and $V \propto R^3$ we find

\begin{equation}
\tau \propto \left(R^2 \nu' \right)^{-(p+4)/2}.
\end{equation}

\section{The case of Nova Muscae}
\label{sec:nm}

\begin{figure*}
\begin{center}
\epsfig{file=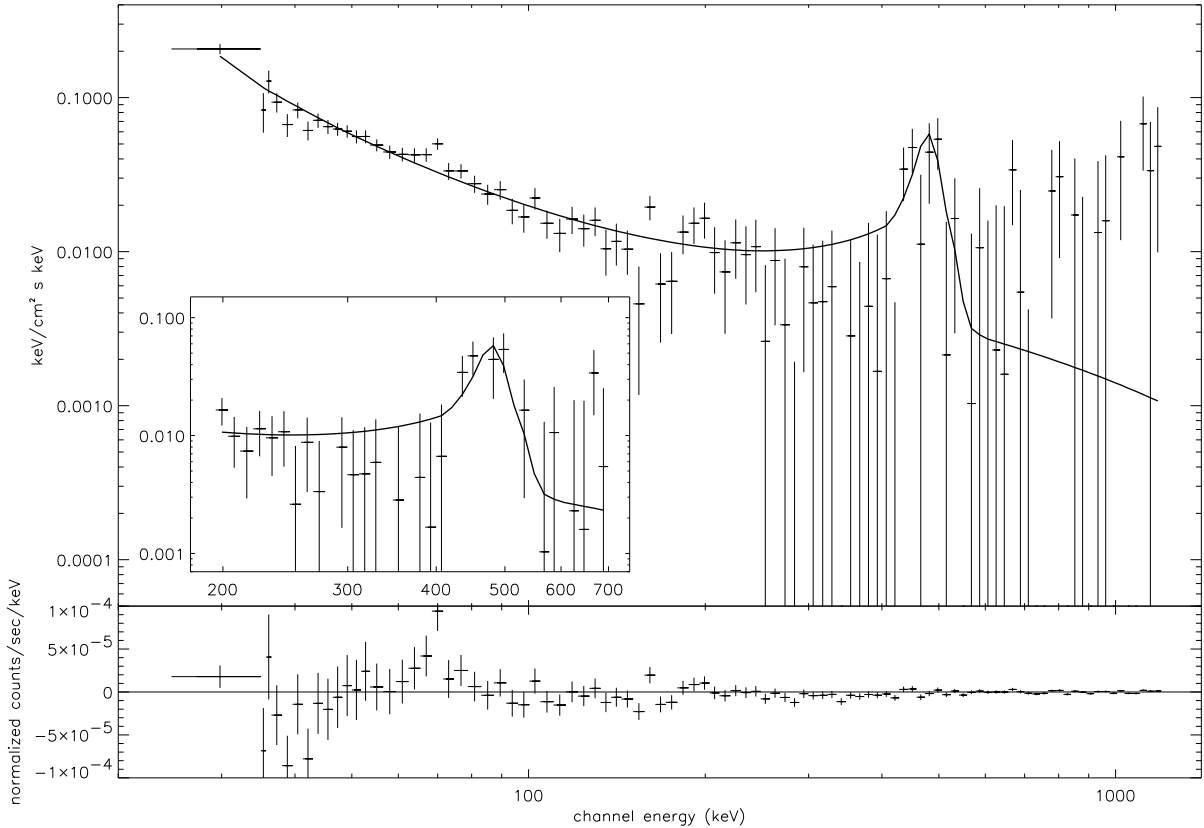, width=16cm}
\caption{The best-fit model to the spectrum of Jan~21.
The large figure shows the full spectrum from the last third of the
Jan~20--21  observation, showing the emergence of the annihilation
features.
The inset  shows the data points around 474~keV
used in the fitting procedure (see the text),  while the bottom
panel shows the  residuals of the model to the data in normalised
counts per second. }
\label{fig:xspec}
\end{center}
\end{figure*}

Nova Muscae (GRS 1124$-$684) was discovered on 1991~January~8 by
Granat (Lund \& Brandt 1991\nocite{lb91}; Sunyaev 1991\nocite{s91})
and Ginga (Makino 1991\nocite{fm91}). An extensive monitoring campaign
with Granat covering most of January and February 1991 followed this
initial discovery (Gil'fanov et al. 1991\nocite{gsc91}, hereafter
GSC91; Sunyaev et al. 1992\nocite{scg92}; Gil'fanov et
al. 1993\nocite{gcs93}). During observations on January~20--21, a
strong, narrow line near 500 keV was detected. This line had not been
noted before and the line flux was observed to increase during the
observation within the space of a few hours. Unfortunately, the
observations stopped before the line flux decreased again, implying a
lifetime of the line emission of at least 10 hours. During subsequent
observations on February~1--2 it was not detected. Simultaneously to
the strong line near 500 keV, there was also increased emission
detected near 200 keV. The spectrum obtained with Granat during
1991~January~20--21 is shown in Figure~\ref{fig:xspec}.  GSC91 fitted
the spectrum with a power-law continuum and two Gaussian line
profiles. The lines peak at $474\pm30$\,keV and $194\pm13$\,keV. The
line fluxes are given as $(6.6\pm3.6) \times 10^{-3}$\,photons
s$^{-1}$ cm$^{-2}$ and $(1.5\pm1.3) \times 10^{-3}$\,photons s$^{-1}$
cm$^{-2}$ respectively. The FWHM of the line centred at 474\,keV is
$70\pm70$\,keV.

\begin{figure}
\begin{center}
\epsfig{file=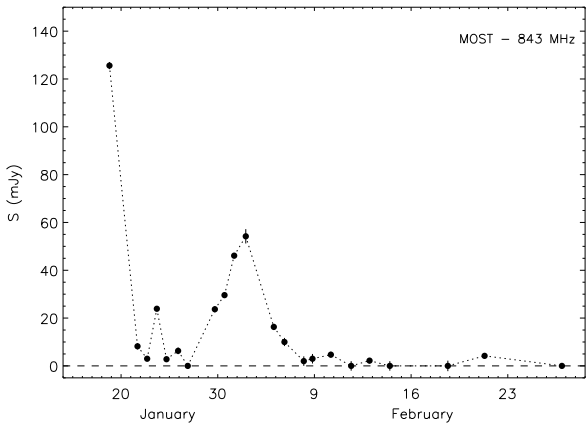, width=8.45cm}
\caption{MOST 843~MHz lightcurve showing the decay of the initial
outburst of January~8 and the two bursts of January~24 and
January~31.}
\label{fig:most}
\end{center}
\end{figure}

The original discovery also triggered a radio monitoring programme
using the Molongolo Observatory Synthesis Telescope (MOST) and the
Australia Telescope Compact Array (ATCA) at frequencies ranging from
843\,MHz to 8.6\,GHz (Kesteven \& Turtle 1991\nocite{kt91}; Ball et
al. 1995\nocite{bkcth95}). The campaign started on January~17 and
continued into early February. The radio lightcurve at 843\,MHz
obtained with the MOST is shown in Figure~\ref{fig:most}. Ball et
al. (1995) note four distinct features in the lightcurve. There is a
general decay of the radio flux from the very beginning of the
programme continuing until about January~22. This is interrupted by a
brief flare around January~18 observed with ATCA (not
shown in Figure~\ref{fig:most}). There is another flare observed from
January~31 until around February~5. Finally, a short flare was
observed by MOST at 843\,MHz on January~24 with a measured flux
density of 24\,mJy. The event lasted only one day.
There were no observations at other frequencies.

In the following we will concentrate on the detection of the
$\gamma$-ray lines on January~21 and the brief radio flare on
January~24.  We speculate that the same ejection event is responsible
for the emission at opposite ends of the electromagnetic spectrum
observed on the two days.

\subsection{Constraints from the annihilation line}
\label{sec:anni}

We assume here that the two $\gamma$-ray lines observed in the
spectrum of Nova Muscae arise from pair annihilation in a relativistic
bipolar outflow from the very centre of the system. In this model, the
line at 474\,keV is associated with the approaching component while
that at 194\,keV arises from the receding component.

Using Equations~(\ref{gamma}) and (\ref{cos}) we then find $\theta =
60^{\circ}\pm7^{\circ}$ for the angle to the line of sight of the
component motion and $\beta = 0.84 \pm 0.02$ for the bulk velocity of
the components. This corresponds to a Lorentz factor of $\gamma _{\rm
b} = 1.86\pm0.12$. The errors are determined from the uncertainties in
the frequency of the line peaks as given by GSC91. With these values
we expect the ratio of the line fluxes of the approaching and the
receding components to be $F_-/F_+=15\pm10$ for discrete components
and $F_-/F_+=6\pm3$ for continuous jets (see
Equation~\ref{fluxratio}).  The observed line fluxes give
$F_-/F_+=4\pm6$. The velocity and angle to the line of sight
calculated from the frequencies of the line peaks are consistent with
this within the errors. However, the uncertainties are very large.

The companion star in Nova Muscae is tidally distorted by the
black hole. The resulting variation in the lightcurve has been
modelled by Orosz et al. (1997)\nocite{obmr96} who constrain the
inclination of the system to $54^{\circ} \le i \le 65^{\circ}$,
and also by Shahbaz, Naylor \& Charles (1997)\nocite{snc97} who
find $i=54^{+20\circ}_{-15}$. It is reasonable to expect that $i$
is also the inclination angle of the accretion disc around the
black hole and that the bipolar outflow moves in a direction
perpendicular to this disc. However, note that the
jets in GRO J1655-40 appear to be misaligned with respect to the
direction perpendicular to the disc (e.g. Orosz \& Baylin 1997).
With these assumptions we should find $\theta = i$, i.e. the angle
of the outflow direction to the line of sight is equal to the
inclination of the system. Our value for $\theta$ is fully
consistent with this interpretation.

We now take a closer look at the profile of the line centred at
474\,keV. The line at 194\,keV is too weak to allow any detailed
investigation of its profile, and hence we only fit the line at
474\,keV for this present study. The FWHM of the line at 474\,keV is
only 70\,keV which implies a pair temperature of only about
$4\times10^7$\,K at the time of annihilation (Ramaty \&
M{\'e}sz{\'a}ros 1981\nocite{rm81}; GSC91).

The rate of annihilation implied by the line flux is $\dot{N}_+
\sim 2\times 10^{43}$\,pairs s$^{-1}$ for a distance of 5.5\,kpc
(Orosz et al. 1996).  This enormous rate is sustained for at least
10 hours (Sunyaev et al. 1992\nocite{scg92}). It is therefore very
unlikely that this feature at 474~keV arises from a large number
of pairs formed practically instantaneously and then slowly
annihilating away. This would imply a short-lived
flash of annihilation photons with a fast, exponential decay
contrary to the lifetime of the annihilation features of at least
10 hours. A more promising approach is to assume that the pair
producing plasma is in equilibrium, i.e. the annihilation losses
are balanced by pair creation. MZC95 failed to fit the line with
such a model assuming a thermal plasma with strong pair escape.
They also tried a model with injection of non-thermal electrons
based on the work of LZ87. Again, the fit was poor. However, they
did not take into account a possible Doppler-redshift of the line.

After correcting the spectrum for the redshift derived above,
we attempted a fit essentially identical to that of MZC95. We
used the model NTEEA within the XSPEC package which is an
implementation of the model developed in LZ87. We disabled any
contribution to the spectrum by reflection off the disc. The
temperature of the black body providing the soft seed photons for the
pair creation process was set to 1\,keV which gives a reasonable fit
to the low energy spectrum of Nova Muscae (see MZC95). Further, we
assumed that there was no thermal heating of the plasma, no escape of
pairs from the source and that the optical depth due to protons was
zero. The energy spectrum of the injected relativistic electrons was
described by $Q_{\rm l} = Q_0 \gamma ^{-2.2}$, extending from $\gamma
_{\rm min}=1$ to $\gamma _{\rm max}=1000$. We have chosen the slope of
the power law to be equal to that of the radio emission (see Section
\ref{sec:radio}). Variation of this parameter as well as changing $\gamma
_{\rm min}$ and $\gamma _{\rm max}$ was found to cause only minor
differences in the result. Finally, the model assumes that the plasma is
contained within a spherical volume. The radius of this we set to
$R=2\times10^5$\,m corresponding to the Schwarzschild radius of a
black hole of mass $6$\,M$_{\odot}$. This is the minimum conceivable
size of the annihilation region. In any case, the exact value of $R$
is only important for the determination of energy losses due to
Coulomb and bremsstrahlung processes. It does not significantly
influence the model fit of the annihilation line. The only free
parameters in this fit were the compactness of the non-thermal
electron injection, $l_{\rm e}$, and the compactness of injected soft
photons, $l_{\rm s}$. The latter is defined to be analogous to $l_{\rm
e}$ with $L_{\rm e}$ in Equation~(\ref{compact}) replaced by the
luminosity of the soft photon injection. As we were only interested in
fitting the annihilation line at 474\,keV, we only used data points in
the range 195--699~keV. We also excluded all negative flux
measurements within this range which resulted in 23 data points to be
fitted.

Figure~\ref{fig:xspec} shows our best fit with a reduced $\chi
^2$-value of 1.35 (for 84~d.o.f). The free parameters,
$l_{\rm e} \sim 3000$ and $l_{\rm s} \sim 50$, imply a strongly
photon-starved plasma. The parameters are not well constrained as
under these conditions the pair yield does not depend strongly on
their exact values (e.g. LZ87). We find a lower limit for $l_{\rm e}$
of about 100, below which the pair yield is too small to give a good
fit. Despite these uncertainties, the correction of the data for a
Doppler-redshift clearly makes a reasonable fit of the annihilation
line possible.

Re-arranging Equation~(\ref{compact}) yields

\begin{equation}
L_{\rm e} = \frac{R m_{\rm e} c^3 l_{\rm e}}{\sigma _{\rm T}}.
\end{equation}

\noindent Substituting in our lower limits for $R$ and $l_{\rm e}$, we
find $L_{\rm e} \ge 7 \times 10^{30}$\,W. From
Equation~(\ref{luminosity}) it follows that $Q_0 = 9\times
10^{26}$\,m$^{-3}$ s$^{-1}$ for this lower limit. The rate at which
relativistic electrons are injected is then $\dot{N}_{\rm inj}=3\times
10^{43}$\,particles s$^{-1}$. This is comparable to the observed
annihilation rate of $2\times10^{43}$\,particles s$^{-1}$.

\subsection{Constraints from the radio emission}
\label{sec:radio}

The radio flare occured on 1991~January~24, about three days after the
observation of the annihilation lines.  If this emission arose from
the same material that was responsible for the annihilation features
at the very centre of the source, then these ejection components were
by then located roughly $7\times 10^{13}$\,m further out.  This
assumes that the components did not decelerate.  When resolved in the
radio, the ejections of X-ray transients are quite collimated
(e.g. Mirabel \& Rodr{\'\i}guez 1994\nocite{mr94}; Hjellming \& Rupen
1995\nocite{hr95}). In order to estimate the size of the
radio-emitting region, we assume a half-opening angle of the ejection
of $1^{\circ}$. This implies a radius of $10^{12}$\,m for a spherical
component. The measured flux density was $23.9$\,mJy at 843\,MHz which
we associate entirely with the approaching component. For the values
found in Section \ref{sec:anni} for the bulk velocity, $\beta =0.84$,
and the angle to the line of sight, $\theta=60^{\circ}$, this converts
to 29.9\,mJy emitted at 909\,MHz in the rest frame of the approaching
component. This assumes a discrete component rather than a continuous
jet. We have no information on the slope of the radio spectrum for 24
January. Therefore, we adopted a spectral index of $\alpha = -0.6$ as
derived from the multi-frequency observations of the earlier
flares. This implies that $P(\gamma) = P_0 \gamma ^{-2.2}$. For a
distance to the source of 5.5\,kpc (Orosz et al. 1996) and using
Equation~(\ref{bmin}), we find the magnetic field strength
correponding to minimum energy conditions to be $B_{\rm min} =
0.015$\,mT. Using Equation~(\ref{wpart}) this implies $W_{\rm part} =
5 \times 10^{32}$\,J and $P_0=8\times 10^8$\, m$^{-3}$. The number of
relativistic particles in the emitting volume is $N_{\rm rel}=2\times
10^{44}$.

Adopting the radius of the emission region at the time of the
observation as a scale radius $R_0=10^{12}$\,m, Equation~(\ref{depth})
for the optical depth at $\nu' =909$\,MHz yields

\begin{equation}
\tau \sim 0.1 \left( \frac{R}{R_0} \right)^{-6.2}.
\label{depth2}
\end{equation}

\subsection{Implications of a bipolar flow}

The annihilation feature observed persists for at least 10 hours
(Sunyaev et al. 1992\nocite{scg92}). The production of an outflow must
have lasted at least for that length of time.  In the case of a
shorter ejection event the annihilation rate would decrease
dramatically as the ejected material travels outwards and expands. Of
course, the annihilation could also proceed in a region close to the
centre of the source without moving outwards. In this case, the
displacement of the line at 474\,keV could be explained as arising
from gravitational redshift. This requires the annihilation region to
be inside about 7 Schwarzschild radii of the 6\,M$_{\odot}$ black
hole. However, in this case it is not clear what causes the second,
simultaneous line at 194\,keV.

An entirely different explanation for the line at 474\,keV is given by
GSC91 and also by Mart{\'\i}n et al. (1994)\nocite{mcmrc94}. This
involves the emission of $\gamma$-ray photons of energy 468\,keV
during the formation of $^7$Li close to the black hole. This is
supported by the observation of optical lithium lines. However, GSC91
point out that the proximity of the lithium production region to the
black hole should result in a strong rotational broadening of the
line. Furthermore, the secondary line at 194\,keV is again not
explained in this scenario.

Misra \& Melia (1993)\nocite{mm93} presented a model for
electron-positron pair jets to explain the annihilation line observed
in the spectrum of the Galactic centre source 1E 1740.7-2942. Their
model is based on a radiative acceleration mechanism for the jets. The
great majority of pairs annihilate within the acceleration zone with a
large variation in the thermal pair temperature. This leads to the
formation of quite a broad line. Also, as mentioned before, a model
with purely thermal pairs is unable to account for the annihilation
line in the case of Nova Muscae (MZC95). In the model presented here
we argue that the bulk of the outflow containing the pairs is already
accelerated when they annihilate, thus explaining the redshifts of the
two observed lines. The velocity of the outflow is then 0.84\,c, while
the radiative acceleration allows only for a terminal velocity of
about 0.7\,c. Because of this, the acceleration of the outflow must,
at least partly, be due to some other mechanism(s).

The energy required to drive the outflows is enormous. As the
relativistic electrons necessary for pair production are highly
relativistic, their mass in the rest frame of the outflow material is
given by $\gamma m_{\rm e}$. Therefore, the total kinetic power of the
relativistic electrons, as measured in the source rest frame, injected
into the outflow can be approximated as

\begin{equation}
\dot{E}_{\rm kin} = V Q_0 m_{\rm e} c^2 \left( \gamma _{\rm b} -1
\right) \int \gamma ^{1-p} \, d\gamma \sim 6\times 10^{31} {\rm \ W}.
\end{equation}

\noindent In this estimate we have used the limiting values for the
electron compactness, $l_{\rm e} = 100$, and the size of the emission
region, $R = 2\times 10^5$\,m. Any increase in these parameters will
also cause the energy estimate to rise. Therefore, the lower limit on
$\dot{E}_{\rm kin}$ corresponds already to about 80\% of the Eddington
luminosity of a 6\,M$_{\odot}$ black hole. Balance of electrical
charge requires the presence of positively charged particles in the
outflow. If these are protons with negligible thermal
energy but travelling at the necessary bulk velocity, then another
$4\times 10^{33}$\,W in kinetic energy are required. This is more than
a factor 50 more than the Eddington luminosity. This energy injection
into the outflow has to be sustained for more than 10 hours and thus
makes a large proton content in the jet very unlikely.

The only alternative is then that from the very beginning of the
ejection the outflow material in the annihilation region consists of
virtually a pure pair plasma.  This removes the requirement of
relatively inefficient pair production from relativistic
electrons. The pairs must be injected into the relativistic bulk flow
with a relativistic velocity distribution to explain the strength of
the annihilation line (MCZ95), the cooling within the outflow and then
annihilation at a rate of $2\times 10^{43}$\,s$^{-1}$. In this case,
the required kinetic power is about $6\times 10^{30}$\,W, or 8\% of
the Eddington luminosity.

Assuming that the outflows consist entirely of pairs, at least
$7\times10^{47}$ pairs are injected into the outflow approaching
us. Most of these annihilate but a small fraction may escape the
annihilation region.  If the radio flare discussed above is associated
with the annihilation line feature and the radio-emitting region only
contains the relativistic pairs required by minimum energy arguments,
then only one in 7000 pairs need escape. In the model of Misra \&
Melia (1993)\nocite{mm93}, roughly 10\% of all pairs escape, which
would imply a large thermal population underlying the radio-emitting
relativistic pairs. Nevertheless, the kinetic energy of this thermal
population is, at $3\times 10^{29}$\,J, small compared to the kinetic
energy of $10^{33}$\,J of the relativistic and therefore heavy
radio-emitting particles. In any case, the energy required to explain
the radio observations is at most only a fraction of 5\% of the
originally injected energy. Thus the radio emission observed in X-ray
transients may represent only an almost negligible fraction of the
energy initially available to the outflow.

The time between the detection of the annihilation line features from
Nova Muscae and the first observations of a radio flare is roughly
three days.  During this time, a number of observations at various
frequencies showed no sign of enhanced radio emission (Ball et
al. 1995\nocite{bkcth95}). This would argue against a connection
between the two events.  However, Equation~(\ref{depth2}) shows that
the outflow became optically thin to emission around 1\,GHz only when
$R\sim 0.7 R_0$. For a constant opening angle of the outflows as
assumed throughout, this implies that radio observations would have
detected this flare only two days after its start at the earliest.

The existence of highly relativistic electrons far out from the centre
of the source implies a secondary acceleration mechanism. The
annihilation line at 474\,keV is consistent with a pair temperature of
only 5\,keV. Therefore the escaping pairs are sub-relativistic and
must be re-accelerated to explain the radio emission. One possibility
is the existence of internal shocks within the outflow (e.g. Kaiser,
Sunyaev \& Spruit 2000\nocite{kss00}).  The duration of the outflow
production is easily long enough to allow for variations in the
outflow speed which may lead to the formation of such
shocks. Alternatively, the particles may be re-accelerated at a
working surface at the end of the flow, as is the case in many
extragalactic jet sources.

\section{Conclusions}
\label{sec:conc}

From the above discussion it is likely that Nova Muscae underwent a
number of bursts resulting in relativistic outflows. The production of
these outflows would then always be accompanied by strong annihilation
line emission followed by radio synchrotron emission a few days
later. Here we are assuming that the radio emission arises in
jets. Unfortunately, only one of the rather minor outbursts of Nova
Muscae in 1991 was covered by both $\gamma$-ray and radio
observations. The strongest outburst leading to the first detection of
the system on 1991~January~8 was observed with insufficient energy
resolution to establish the presence of an annihilation
line. Subsequent high-resolution observations on January~9 did not
detect any lines. There were two more minor radio bursts. One of them
occurred on January~18 and a longer burst started around January~31
(Ball et al. 1995\nocite{bkcth95}). If the three days' delay between
the appearance of the annihilation lines and the start of the radio
emission is typical, then the two radio flares should have been
preceeded by annihilation lines on January~15 and around January~28
respectively. There were no $\gamma$-ray observations during the
period from January~11 until January~16 and from January~22 until
February~1 (GSC91).  The non-detections of January~9 and January~16
may be taken as evidence that the outflow production, when it occurs,
is rather short-lived.

If it is true that the annihilation features observed originate from
accelerated plasma, then protons as a contributing factor to the
matter in the plasma are ruled out. The scenario we have discussed
favors a pair plasma --- even if only a tiny fraction of the pairs
survives, this is sufficient to produce the observed radio emission.
Furthermore, the detection of Doppler-shifted annihilation lines is a
direct method, independent of uncertainties arising from other
observational parameters such as fluxes, by which to directly measure
jet speeds.

However, to confirm the outflow scenario linking $\gamma$-ray line
observations with radio emission, further detailed, time-resolved
observations are needed. The spectral and temporal resolution of {\sc
integral} will enable the detection and follow the evolution of
possible annihilation features in other jet sources. This will prove
extremely valuable in the case of, for example, GRS~1915$+$105 which
is heavily obscured at visible wavelengths.  In addition, this study
was applied to Nova Muscae which is a transient, and hence it will be
interesting to observe the behaviour of permanent jet sources, such as
1E~1740.7$-$2942, from which an annihilation line has already been
detected.

\section*{Acknowledgements}

We would like to thank Marat Gil'fanov for providing us with the
original {\sc GRANAT} data. It is also a pleasure to thank Andrzej
Zdziarski for help with {\sc NTEEA}.  We also thank Duncan
Campbell-Wilson and Lewis Ball for so generously handing us the MOST
and ATCA data for this project. The MOST is operated by the University
of Sdyney and funded by grants from the Australian Research Council.
We thank the anonymous referee for helpful comments. DCH acknowledges
the support of a PPARC postdoctoral research grant to the University
of Southampton.

%\bibliography{../../crk}
%\bibliographystyle{../../styles/mnras}

\end{document}